\documentstyle[epsfig]{mn2e}

\begin{document}
\title[Disc response to vertical perturbations]
{Response of a galactic disc to vertical perturbations : Strong dependence on 
density distribution}
\author[P. Pranav and C.J. Jog ]
       {Pratyush Pranav$^{1}$\thanks{E-mail : pratyuze@physics.iisc.ernet.in}, and
        Chanda J. Jog$^{1}$\thanks{E-mail : cjjog@physics.iisc.ernet.in}\\
$^1$   Department of Physics,
Indian Institute of Science, Bangalore 560012, India \\
}

\maketitle

\begin{abstract} 
We study the self-consistent, linear response of a galactic disc to non-axisymmetric
perturbations in the vertical direction as due to a tidal encounter, and show that the density distribution near the 
disc mid-plane has a strong impact 
 on the radius beyond which distortions like warps develop. 
The self-gravity of the disc resists distortion in the inner parts.
Applying this approach to a galactic disc with an exponential vertical profile, Saha \& Jog showed that warps develop beyond 4-6 disc scalelengths, which could hence be only seen in HI.
The real galactic discs, however, have less steep vertical density distributions 
that lie between a $sech$ and an exponential profile.
Here we calculate the disc response for such
a general $sech^{2/n}$ density distribution, and show that the warps develop from a smaller radius of  2-4 disc scalelengths. This naturally
explains why most galaxies show stellar warps that start within the optical radius.
Thus a qualitatively different picture of ubiquitous optical warps emerges for the observed less steep density profiles.
The surprisingly strong dependence on the density profile is due to the fact that 
the disc self-gravity depends crucially on its mass distribution close to the mid-plane.
General results for the radius of onset of warps, obtained as a function of the disc scalelength and the vertical
scaleheight, are presented as contour plots which can be applied to any galaxy.

\end{abstract}


\section{Introduction} 

It is a common knowledge now that the galaxies are not isolated structures in
the universe, rather galaxy interactions including mergers are common.
Hence understanding the dynamics of these interactions and the effects they produce
on the structure and evolution within galaxies has been of much interest. 
One such effect is the generation of 
asymmetric features due to tidal encounters between galaxies.

Spiral galaxies are known to display a variety of non-axisymmetric features, both in the plane  and also in the direction perpendicular to the plane. The most common vertical distortion is a warp, a feature of azimuthal
wavenumber $m=1$. Most nearby galaxies show  such integral-sign or s-shaped warps in their outer parts (e.g., Binney \& Tremaine 1987). A similar planar distortion commonly seen is the lopsidedness
in disc corresponding to azimuthal wavenumber $m=1$ (Jog \& Combes 2009). 
The origin of warp is not yet clear despite a long search.
 A commonly suggested mechanism for the origin of warps is due to the tidal interaction with its neighbours (e.g.,
Schwartz 1985, Zaritsky \& Rix 1997). Weinberg (1995)  studied the generation of warp in our
Galaxy due to perturbation from the 
neighbouring Large Magellanic Cloud.

Vertical distortions other than warps are also commonly observed in spiral galaxies.
Small-scale corrugations in external galaxies have been studied (Quiroga 1984) and so have
been the distortions in the stellar distributions of galaxies (Florido et. al
1991). Scalloping in HI in the outer regions of our Galaxy has been studied too (Kulkarni, Blitz \& Heiles
1982). Recently Matthews \& Uson (2008)  find that corrugation in HI is seen in IC2233 even in the inner regions.

While warps are seen mostly in the outer parts of a galactic disc, 
surprisingly little work has been done to discuss the radius at which warps develop.
Saha \& Jog (2006) proposed this as being determined due to the self-consistent disc response to a
tidal field. The disc self-gravity resists distortion in the inner parts, and only in the outer parts does the disc begin to respond to the external potential.

In this paper, we continue with this approach and study self-consistent vertical distortions for different forms of vertical density distributions  in the disc. We are motivated by
the fact that in the previous studies, not much attention has been paid to the
effect the form of density distribution for the responding disc might have on
the overall behaviour of the system. For the purpose of modeling, the
vertical distribution was taken to be exponential for
mathematical simplicity in an earlier work (Saha \& Jog 2006).

The study of vertical distribution of stars in galactic discs has an interesting history.
The vertical distribution 
was at first deduced to be of type 
$sech^2$ as resulting for an isothermal disc (Spitzer 1942). However,  observations showed a steeper profile closer to a $sech$ or an exponential both for our Galaxy (e.g., Gilmore \& Reid 1983, Kent et al. 1991) as well as for external galaxies (e.g., Wainscoat, Freeman \& Hyland 1989, Rice et al. 1996).  An exponential profile all the way to the mid-plane is unphysical and hence van der Kruit (1988) proposed a generalized function $sech^{2/n}$ to represent the vertical density distribution. In this scheme, $n=1$ and 2 correspond to a $sech^2$ and a $sech$ distribution and as $n$ tends to $\infty$ it asymptotically approaches an exponential distribution. Later observers analyzed their data and cast in this format, and have shown that a true density distribution is less steep than an exponential
 and in most case lies between a $sech$ and an exponential distribution (Barteldrees \& Dettmar 1994, 
de Grijs, Peletier \& van der Kruit 1997).
While dust extinction prevents a determination of the stellar density profile close to the mid-plane, the near-infrared bands do not have this limitation and represent the true density profiles  representing the old stars. For a sample of 24 galaxies studied in the K$_s$ band,  de Grijs et al. (1997) find that, a mean value of $<2/n> = 0.5$ corresponding to the  $n$ index = 4. Thus the vertical density profile for stars in a typical  galactic disc lies between a $sech$ and an exponential profile, being closer to a $sech$.

In this paper, we study the vertical response of an axisymmetric disc to an external imposed
perturbation, where the disc density follows such a general $sech^{2/n}$  
 distribution. We also study the response of an exponential disc for the sake
 of comparison. 
The motivation for our work comes from the fact that the matter distribution
close to the mid-plane contributes strongly to the  vertical
self-gravitational force (Banerjee \& Jog 2007), hence it is plausible that
the different vertical density profiles affect the disc response in
different ways. We
show that indeed the disc response has a strong dependence on 
the vertical density distribution. This in turn significantly affects the radius for the onset of
various non-axisymmetric features along the vertical direction.

Section 2 of the paper presents the details of the model and the methods of
calculation, while section 3 presents the results. Section 4 presents the
discussion and section 5 concludes the paper.

\section{Details of the model}

\subsection{Density response of the perturbed disc}

In this paper, we study the linear response of the disc to an external imposed perturbation
potential for a general $sech^{(2/n)}$ vertical distribution, as well as the exponential case for comparison (see Section 1). In an earlier work, the density distribution perpendicular to the plane of the disc was taken to be
 exponential for simplicity (Saha \& Jog 2006).
 Here we build on the formulation developed in that earlier paper,
with the main difference being that the density profile $\rho (z)$ is taken to be a general, flat-core of type
 $sech^{(2/n)}$ in this case.
Also, the radial variation of the potential is taken into account properly here, which leads to 
a more realistic behaviour of the response function.

We use cylindrical co-ordinates. The stellar density distribution of the unperturbed axisymmetric disc is taken as:
\begin{equation}
\rho_d(R,z) =  f_1(R) f_2(z)
\end{equation}
where $f_1(R)$ is a function corresponding to the radial dependence and $f_2(z)$ is a
function which takes care of the dependence in the vertical direction. The radial dependence is taken
to be exponential (Freeman 1970), while  a $sech^{(2/n)}$ dependence in the vertical
direction is taken, as based on the work by van der Kruit (1988). However, we want to consider the case when the mid-pane density at a given radius is same for any $n$ value, hence we use the form as in Kuijken \& Gilmore (1989, see eq.[A13]) :
\begin{equation}
f_1 (R) = \rho_{d0} \: exp ^{(-R/R_D)}; \: \: f_2 (z) =    sech^{2/n} (\frac{nz}{2z_0})
\end{equation}
\noindent where $\rho_{d0}$ is the central, mid-plane density and $R_D$ is the exponential disc scalelength. 
Note that for $n =1,2 $, the function $f_2(z)$ corresponds to a $sech^2$ and a $sech$ profile respectively, while
for $n {\rightarrow \infty}$, it reduces to a form
exp $(- z/ z_0)$. Thus z$_0$ is taken to denote the vertical scaleheight.
Hence,
\begin{equation}
\rho_d(R,z)=\rho_{d0} \: exp^{(-R/R_{d})} \:   sech^{2/n} (\frac{nz}{2z_0})
\end{equation}
In order to see the effect that the halo dark matter distribution might have on the
response, for the sake of completeness, we also include an axisymmetric spheroidal system with a
pseudo-isothermal density profile (de Zeeuw \& Pfenniger 1988):
\begin{equation}
\rho_h(R,z)=\frac{\rho_{h0}}{1 + [R^2+(z^2/q^2)] / R_c^2}
\end{equation}
\noindent where $\rho_{h0}$, $R_c$ and $q$ are the central density, the core radius, and
the flattening parameter or the vertical to planar axes ratio respectively. 
 The halo symmetry plane (z=0) is
assumed to coincide with the disc plane. 

The density distributions of the halo and the disc are related to the
corresponding potentials by the Poisson equation 
\begin{equation}
 \frac {1}{R} \frac{\partial}{\partial R} \large (R \frac{\partial \Psi}{\partial R} \large ) =  4 \pi G \rho
\end{equation}
\noindent where $G$ is the gravitational constant and the total density $\rho$ is: 
\begin{equation}
\rho=\rho_{d}+\rho_{h}
\end{equation}
For a ``disc plus halo'' system, the total potential $\Psi$ is the
combination of the individual potentials of the disc and halo
\begin{equation}
\Psi = \Psi_{d}+\Psi_{h}
\end{equation}

For a vertical distribution as in eq. (2),
 the corresponding integral obtained after inverting the Poisson equation is not
solvable analytically. However, it can still be represented as a sum of integrals
which converge fairly rapidly:
\begin{eqnarray}
\Psi_d(R,z)  =  -2\pi G \int_0^{\infty}\,dk J_0(kR)\rho_d \alpha(\alpha^2+k^2)^{-3/2}2^{1+(2/n)}\nonumber\\
\times \sum_{m=0}^{\infty} {{-2/n} \choose m}
\frac{(1/z_0)(1+ n m)e^{-k|z|}-ke^{(1+ n m) {z/z_0}}}{(1+ n m)^2 (1/z_0)^2-k^2}\nonumber\\
\end{eqnarray}
\noindent where $\alpha = 1/R_D$, and $J_0 (kR)$ is the cylindrical Bessel function of the first kind, of order zero.  The $m^{th}$ binomial coefficient of power -(2/n) is denoted by
${-2/n} \choose {m}$ (Sackett \& Sparke 1990). If it so happens that n, k and
$z_0$ have values such that the denominator has a zero value, such terms can
still be evaluated by removing the singularity using L' hospital's Rule. The
resulting term is then
$\frac{2^{2/n}}{k}$ ${-2/n} \choose {m}$ $(1+k|z|)e^{-k|z|(2/n)}$.

The potential corresponding to the halo density distribution is not
analytically obtainable either, but following the more general expression for
obtaining potential corresponding to any distribution, we can reduce it down to
a form which is numerically integrable (Binney \& Tremaine 1987, eq. 2.88). We have, then
\begin{eqnarray}
& & \Psi_h(R,z) = 2 \pi G \rho_{ho} q {R_c}^2 \nonumber\\
& &  \: \: \: \: \: \: \: \: \times \int_0^{1/q}ln[1+\frac{x^2}{{R_c}^2}(\frac{R^2}{\epsilon^2x^2+1}+z^2)]dx/(\epsilon^2x^2+1)\nonumber\\ 
\end{eqnarray}
\noindent where $\epsilon =(1-q^2)^{1/2}$ is the eccentricity of the halo.

We consider the zero-forcing frequency perturbation potential due to a
perturber situated at a distance D from the centre and at an angle of
inclination i from the disc mid-plane to be:
\begin{equation}
\Psi_1(R,\phi,z)=\Psi_{p0}(R)(1+\frac{z}{D}\sin i)cos(m\phi)
\end{equation}
In the limit that $R/D\ll1$ and $z/D\ll1$, $\Psi_{p0}={GM_pR}/{D^2}$, is a
slowly varying function of the radius. 
We treat this more realistic case with the effect of perturbation increasing with radius.
In the earlier study (Saha \& Jog 2006) this was taken to be a constant.

The total potential that a particle in the disc experiences is a
superposition of the potential generated by the disc plus the dark matter halo system and the
external perturber, given by equations (7) and (10) respectively:
\begin{equation}
\Psi_{total}=\Psi(R,z)+\Psi_1(R,\phi,z)
\end{equation}
To obtain the vertical response of a disc to an imposed perturbation, one needs to solve the equation
of motion as a forced oscillator, and combine with the continuity equation, and with the Poisson equation for the perturbation. See Saha \& Jog (2006), section 2.1 for details.  This gives the following for the perturbed motion of a particle:
\begin{equation}
z_m = H (R) cos (m {\Omega} t)
\end{equation}
\noindent where
\begin{equation}
H(R) = \frac {\frac {\partial \Psi_1}{\partial z} \vert _{(R,0)} }{{{\nu}_0}^2 - {{\Omega}_0}^2}
\end{equation}
\noindent Here ${\nu}_0$ and ${\Omega}_0$ are the vertical and planar frequencies of oscillation in the unperturbed system
and are defined respectively as:
\begin{equation}
{\nu_0}^2 = \frac {{\partial}^2 \Psi}{\partial z^2} \vert_{(R,0)} \:\: ; \: \: {\Omega_0}^2 = \frac{1}{R_0} 
\frac{\partial \Psi}{\partial R}\vert_R
\end{equation}
The perturbed mass density $\rho_1$ is given by:
\begin{equation}
\rho_1 (R, \phi, z) = - \frac{\partial \rho}{\partial z}  H(R_0)  cos (m \phi) 
\end{equation}
\noindent where $\rho$ is the net unperturbed density = $\rho_d + \rho_h$ (eg.[6]). Using the equation (2) for the vertical density distribution in the disc, and noting that
 $d \rho_d / dz = - [\rho_d \: tanh (n z/ 2 z_0)] \: / \: z_0 $, the perturbed density reduces to:
\begin{eqnarray}
\rho_1 (R, \phi, z) =  [\rho_d (R, z) tan h (n z /2 z_0) + \zeta_h (R,h) ] \large [ \frac{H(R)}{z_0} \large]\nonumber\\
\times  cos (m \phi) \nonumber\\
\end{eqnarray}   
\noindent On simplifying  $H(R)$ using eqs. (10) and (13), this reduces to:
\begin{equation}
\rho_1 (R, \phi, z) = \rho_r (R, z) \: cos (m \phi) 
\end{equation}
\noindent where $\rho_r(R, z)$,
the magnitude of the perturbed density, is:
\begin{eqnarray}
& & \rho_r(R, z)=\frac {(G M_p R/ D^3){sin i}}{{\nu_0}^2-{\Omega_0}^2}\nonumber\\
& &  \: \: \: \: \: \: \: \: \times \frac{\large[\rho_d(R, z) tan h(n z/2z_0)+\zeta_h(R,h)\large]}{z_0}\nonumber\\
\end{eqnarray}
Here $\zeta_h (R, h)$ (eq.[16]) is defined to be
 = $- z_0 {\partial \rho_h}/{\partial} z$, note that this
is a positive definite quantity for any realistic, centrally concentrated
gravitating system. Further, since the perturbation potential has a positive gradient ($\partial \Psi_1 / \partial z > 0)$, and ${\nu_0}^2$ the vertical frequency square is greater than ${\Omega_0}^2$, the planar frequency square in an
oblate geometry, hence the function $H(R) > 0$. Thus the perturbed disc response follows the perturbing potential. This 
has important implications since the corresponding response potential opposes the imposed potential, as discussed
in the next section. This concept was first noted and discussed for the planar perturbation such as lopsidedness, by Jog (1999). 

Most studies of galactic structure assume a spherical or an oblate halo for which the above relation 
(${\nu_0}^2 > {\Omega_0}^2$) is valid (see e.g., the model for our Galaxy by Mera, Chabrier \& Schaeffer 1998). A prolate halo would not permit this, especially in the outer parts where the halo dominates but such a halo is not indicated. Recent studies of halo shape obtained by modeling the observed HI scaleheights have shown that
the halo is spherical as in our Galaxy (Narayan, Saha \& Jog 2005, Kalberla et al. 2007),  and in UGC 7321 (Banerjee, Matthews, \& Jog 2010), or is oblate as in M31 (Banerjee \& Jog 2008).  
Here we study a nearly spherical halo, with a small oblateness ($q$, the axis ratio = 0.95).

\subsection{Disc response potential}

We next obtain the
 potential corresponding to the density response of the disc to the imposed potential (eq.[16]). This is obtained by inverting the Poisson equation. The Poisson equation is given by:
\begin{equation}
{\nabla}^2 {\Psi}_{resp} (R, \phi, z) = 4 \pi G {\rho_1 (R, \phi, z) }
\end{equation}

 For a disc of non-zero thickness, we solve the Poisson equation by using the Green's function approach (Jackson 1975), see Saha \& Jog (2006) for details.  
This is an integral function approach which gives a numerically tractable form of the resulting response potential.
The Green's function in a cylindrical form for a finite thickness is developed following the treatment in Binney \& Tremaine (1987, see chapter 2).
\begin{equation}
\Psi_{resp}(R,\phi,z)=-G\int_{\emph{V}}\frac{\rho_1(\vec{r}')}{|\vec{r}-\vec{r'}|}d^3\vec{r}'
\end{equation} 
where $V$ denotes the volume of the galaxy, and
\begin{equation}
|\vec{r}-\vec{r}'|=[R^2+R'^2-2RR'\cos(\phi'-\phi)+(z-z')^2]^{1/2}
\end{equation}
Using the above, the equation for response potential for the Fourier component $m$ becomes
\begin{eqnarray}
& & \Psi_{resp}^m (R,\phi,z) = - G  \nonumber \\
& & \times \int_{\emph{V}}\frac{\rho_r(R',z')cos(m\phi')R'\,dR'd\phi'dz'}{[R^2+R'^2-2RR'\cos(\phi'-\phi)+(z-z')^2]^{1/2}}  \nonumber \\
\end{eqnarray}
\noindent where $\rho_r (R,z)$ is the magnitude of the density response (see eq.[18]). The denominator in the above equation can be simplified as:
\begin{eqnarray}
& &[R^2+R'^2-2RR'\cos(\phi'-\phi)+(z-z')^2]^{1/2}\nonumber\\
& & = [(R + R')^2+(z-z')^2]^{1/2} \times [1-k^2 cos^2(m[\phi-{\phi}']/2)]^{1/2}\nonumber\\  
\end{eqnarray}
\noindent where $k^2$ is defined as
\begin{equation}
k^2 = \frac { 4 R R'} { (R + R')^2 + (z - z')^2 + {z_s}^2 }
\end{equation}
Here $z_s$ is the small softening parameter added to prevent the response potential from blowing up at $z = z'$.

Next, eq.(22) is simplified by substituting from eq.[23] into it:
\begin{eqnarray}
\Psi_{resp}^m(R,\phi,z)= - G{\int_0}^{\infty}{\int_{- \infty}}^{\infty}\frac{\rho_r (R,z) R'dR'dz'}{[(R+R')^2+(z-z')^2]^{1/2}} \nonumber\\
\times {\int_0}^{2 \pi}\frac{{cos m {\phi}'} d {\phi}'}{[1 - k^2 cos^2 (m [\phi - {\phi}']/2)]^{1/2}} \nonumber\\
\end{eqnarray} 
\noindent In the above equation, the integral over $\phi$ can be solved by
introducing a variable $\beta$ defined as
\begin{equation}
\beta = (\phi - \phi')/2 \: + \pi /2
\end{equation}
The integral over $\phi$ then reduces to:
\begin{eqnarray}
& & {\int_0}^{2 \pi}\frac{{cos m {\phi}'} d {\phi}'}{[1 - k^2 cos^2 (m [\phi - {\phi}']/2)]^{1/2}} \nonumber\\
& &  = \pm 4 cos m \phi \: \times {\int_0}^{2 \pi}\frac{{cos 2 m {\beta}} d {\beta}}{ 1 - k^2 sin^2 {\beta} } \nonumber\\
\end{eqnarray} 
\noindent The negative sign is valid for odd values of $m$= 1,3,5 etc; while 
for even values of $m =2,4,6$ etc the integral has a positive sign. 
The integral over $\beta$ can be recast in terms of the more familiar elliptical integrals. On doing this, and substituting back in eq.[25], we get the net response potential for odd values of $m$ to be as follows:
\begin{eqnarray}
& & \Psi_{resp}^{odd}(R,\phi,z,m) = -4G\cos(m\phi) \nonumber\\
& & \times\int_0^{\infty}\int_{-\infty}^{\infty}\frac{\rho_r(R',z')R'dR'dz'}{[(R+R')^2+(z'-z)^2]^{1/2}}[2\Theta_{odd}(m,k)-\emph{K}(k)] \nonumber\\
\end{eqnarray}
where
\begin{equation}
\Theta_{odd}(m,k)=\int_0^{\pi/2}\frac{sin^2(m\beta)}{[1-k^2sin^2(\beta)]^{1/2}}d\beta
\end{equation}
and $\emph{K}(k)$ is the complete elliptical integral of the first kind defined as:
\begin{equation}
K (k) =  \int_0^{\pi/2}\frac{1}{[1-k^2sin^2(\beta)]^{1/2}} d\beta
\end{equation}
\noindent where $k$ is as defined in eq.(24).

In case of even $m$ values, the equation retains a  similar form except we have, instead of $\Theta_{odd}$, a similar
function $\Theta_{even}$ defined as
\begin{equation}
\Theta_{even}=\int_0^{\pi/2}\frac{\cos^2(m\beta)}{(1-k^2sin^2\beta)^{1/2}}d\beta
\end{equation}
This in turn gives us the equation for the disc response potential for the
even $\emph{m}$ values as:
\begin{eqnarray}
& &\Psi_{resp}^{even}(R,\phi,z,m) = + 4 G \cos(m\phi) \nonumber \\
& &
\times \int_0^{\infty}\int_{-\infty}^{\infty}\frac{\rho_r(R',z')R'dR'dz'}{[(R+R')^2+(z'-z)^2]^{1/2}}
[2\Theta_{even}(m,k)-\emph{K}(k)]  \nonumber \\
\end{eqnarray}

In equations (28) and (32),  the quantity $\rho_r$ is equal to the magnitude of the response density as given by eq.[18].

From here on,
our treatment differs from that of Saha \& Jog (2006), 
in that we do not take the perturbation potential, $ \Psi_{p0} = GM_pR/D^2$ outside the integral over $R$ (which was not justified since R $\sim R'$); and we do not ignore the contribution of the halo term within the integral over radius. 
 These mathematical refinements, in particular the correct treatment of the radial variation in the numerical calculations, lead to substantially different results at large radii and for high $m$ values compared to Saha \& Jog (2006) even for the exponential case treated in that paper, see Section 4.

We obtain the response potential numerically, and find that it has a sign opposite to the 
imposed potential (eq.[10]), thus it is negative.  Thus the potential corresponding to the disc response to the 
imposed potential alone opposes it. This is a general result for any self-gravitating system, and was first shown for 
the planar case for $m = 1$ (Jog 1999) and $ m = 2$ and $3$ (Jog 2000). 
{\it Thus the negative disc response is a general feature applicable for any self-gravitating system subjected to an external perturbation.}

We next define a dimensionless response potential $\eta_m$ for an azimuthal component $m$, at z=0 as:
\begin{equation}
\eta_m=\frac{|\Psi_{resp}^m|}{\Psi_1}
\end{equation}
\noindent where the response potential $\Psi_{resp}^m$ is given respectively by equations (28) and (32) for odd and even $m$ values, and $\Psi_1$ is the perturbation potential (eq. 10), all defined at the mid-plane, z=0. 
Note that both the numerator (via the term $H(R,z')$) and denominator are proportional to the term $G M_p /D^2$ in the perturbing potential. Hence the ratio $\eta_m$ is independent of the strength of the linear perturbation potential.

\subsection {Self-consistent disc response}

So far we have treated the disc response to the imposed perturbation potential alone. However the disc gravity will also play a role in determining the net disc response. We next obtain the net, self-consistent disc response, following the approach of Jog (1999, 2000). A particle in the disc will be affected by both the imposed potential, $(\psi)_m$, and also the potential corresponding to the disc response to it. For a self-consistent case, the net potential, $(\psi_{net})_m$, for the azimuthal number $m$  can be written as:
\begin{equation}
(\Psi_{net})_m = (\Psi_1)_m + {(\Psi_{resp})'}_m   
\end{equation}
\noindent where ${(\Psi_{resp})'}_m$ is the disc self-gravitational potential which corresponds to the disc response to the net potential, for the Fourier component $m$. In analogy
 with the direct disc response to the imposed potential alone (eq.[33]), the above can be written as:

\begin{equation}
{(\Psi_{resp})'}_m = - \eta_m  (\Psi_{net})_m 
\end{equation}

Substituting this in the previous equation, we get:
\begin{equation}
(\Psi_{net})_m  = \frac {(\Psi_1)_m} {1 + \eta_m} = (\Psi_1)_m  \delta_m 
\end{equation}
\noindent where $\delta_m (\leq 1) $ is defined to be the {\it reduction factor} for the $m^{th}$ Fourier component, and denotes the fraction by which the magnitude of the imposed external potential is reduced due to the self-consistent negative disc response. Thus,

\begin{equation}
\delta_m = \frac {1}{1+\eta_m}
\end{equation}

Note that
$\delta_m$ is always less than or equal to one by virtue of its definition.
It tells us  how strongly the disc reacts to
counteract the effect of an external perturbation. Lower values of $\delta$
denote a higher resistance due to self-gravity to the external perturbation. In turn, the regions
with high $\delta$ indicate that the self gravity is weaker there and the disc is more
susceptible to influences from the external field. 
From eq. (37), we can see that the limiting value of
$\delta=1$  corresponding to $\eta=0$ denotes 
that the disc-gravity plays no role
in this case and it can be taken to be exposed directly to the external
potential. On the other hand, $\delta < 1$, indicates a reduction in the net potential
due to disc self-gravity. 
The net potential determines the perturbed motion that is actually seen,  which results in  a warp for the case $m=1$.

Although the negative disc response may seem somewhat surprising at a first glance, in reality it is a general feature, and we expect this to be seen in any gravitating system that is perturbed by an external field. 
%It is as if the 
It shows that the core of the self-gravitating system is 
left undisturbed while the outer regions suffer the consequences of the external perturbation potential.

\section{Results}

We study the radial dependence of the resulting response potential for the different Fourier components $m$, for various vertical density distribution profiles in
the disc. 
We obtain the plots of the reduction factor vs. radius and see where the minimum lies. This is next argued to denote the 
location of the onset of that particular vertical distortion in the disc.

The perturbed motion is still described by the equation of motion (see eqs. [12],[13]), except that now the perturbation potential is replaced by the net potential (eq.[36]) that takes account of the self-consistent disc response.
Thus the reduction factor has a minimum at say, R$_{min}$, while $\psi_1$ the imposed perturbation (eq.[10]) increases linearly with radius. Hence the net potential has a minimum around this radius, and the net vertical disturbance will be seen mainly in the outer parts beyond this radius. 
Hence, we define
R$_{min}$ to be the {\it radius of onset of vertical distortion} for the case $m$, with $m=1$ for warps, as done in Saha \& Jog (2006). The warp amplitude thus increases with radius beyond the onset radius.
This notion of the onset radius is particularly applicable if the reduction factor has a sharp minimum.

The magnitude of $\delta$ at R$_{min}$ and beyond is also an important result since it tells us how strongly the disc responds to the external perturbation at a given radius, in view of the reduction due to the disc self gravity.

\subsection {Input parameters, and numerical solution}

The input parameters for the disc are:  $\rho_{d0}$, the disc central density; $R_D$, the disc scale-length;  and  $z_0$, the vertical scalelength for the $sech^{2/n}$ function used.
The vertical density profile at any radius is denoted by the index $n$ (see eq. [2]), where $n=$ 1, 2 and $\infty$ correspond to a $sech^2$, $sech$, and an exponential vertical profile respectively.
The halo parameters are: 
$\rho_{h0}$, $R_c$ and $q$, or the central density, the core radius and
the flattening parameter respectively.

We consider a typical spiral galaxy with disc and halo parameters as for our Galaxy, 
from the model by Mera et al. (1998). Thus, the central surface density is 
640 M$_{\odot}$ pc$^{-2}$ (see Narayan \& Jog 2002), and the vertical scaleheight for an exponential distribution is  300 pc, hence we take the central mid-plane density $\rho_{d0}$ to be = 1 M$_{\odot}$ pc$^{-3}$. The disc scalelength R$_D$ is taken to be 3 kpc, since this also agrees with that for a typical giant spiral (Binney \& Tremaine 1987).
Thus, the ratio of vertical scale-height and exponential disc
scale-length is taken to be a constant at
 $z_0/R_d = 0.10$. Later these values are varied to consider the dependence on these.

From the Mera et al. model (1998), the halo parameters are taken to be:
$\rho_{h0}$ = 0.035 M$_{\odot}$ pc$^{-3}$, and $R_c$ = 5 kpc. The halo is taken to be nearly spherical with a small oblateness, with the
flattening parameter  q = 0.95 (see Section 2.1). Although the dark matter halo is included for the sake of completeness, it has a negligible effect on the disc response, since the halo is less concentrated near the mid-plane than the disc is
(eq.[18]).  This is true even for a flattened halo as seen in cosmological simulations, see the Appendix A for details.

The perturber is taken to be a satellite like
the Large
Magellanic Cloud(LMC) situated at a distance $D = 48.5 kpc$, and at an angle $i=90^0$ for simplicity. 
The parameter $M_p$, the mass of the perturber, is taken to be
$2X10^{10} M_{\odot}$ (eq.[10]).
In the current work, we also systematically calculate the values of the reduction factor, $\delta_m$, for $m
= 1$ to $10$  for a $sec h$ vertical distribution. 

Equations (28) and (32) are solved numerically using adaptive quadrature method for radial
points spaced at 0.1 kpc. We go upto a radial distance of 60 kpc in our
calculations for the sake of completeness. This value is beyond the observed extent of usual
galactic discs is known to be.

\subsection {Dependence on vertical density distribution}
Figure 1 contains a plot of the resulting reduction factor, $\delta_1$ for $m=1$ for the various
typical disc vertical density distributions considered (from eq.[2]).
\begin{figure}
\centering
\includegraphics[width=2.9in,height=2.4in]{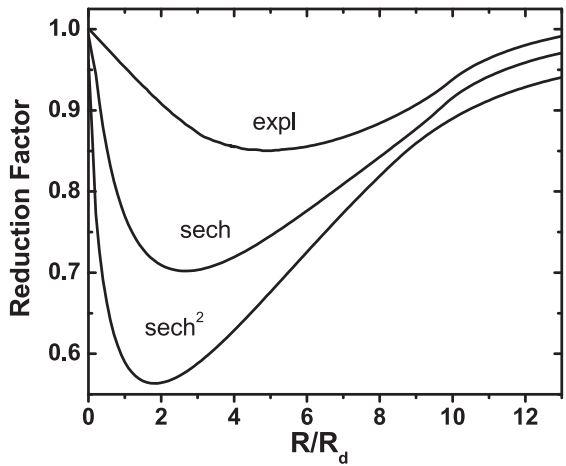} 
\caption{Plot of $\delta_1$, the reduction factor for $m=1$ vs. radius $R/ R_D$, for different vertical 
density profiles ($sech^2$, $sech$ and an exponential) in the galactic disc. 
As the disc profile becomes more flat towards the mid-plane  as in the case of $sech$ and $sech^2$, the disc self-gravity 
is  more important. Thus the disc resists distortion resulting in a lower reduction factor. In contrast, 
for the exponential case, the reduction factor is higher, hence the disc is more readily responsive to the external potential.}
\end{figure}
The plot shows that the minimum in the reduction factor and its location depend strongly on the choice of the vertical density distribution function. The  explanation for this lies in the
difference that the exponential and $sech^{2/n}$ distributions produce. 
The $sech^2$ profile shows the lowest value of the minimum of the reduction factor, indicating the strongest negative disc response due to self-gravity. This is due to a higher mass concentration closer to the mid-plane. In this case
the distribution is much flatter and smoothly continuous near the mid-plane, 
thereby making it more massive close to the mid-plane and hence more resistant
to being influenced by an external disturbance.
On the other hand, an exponential distribution is characterized by a sharp cusp near the mid-plane
denoting less amount of matter and hence lower self-gravity. Hence the disc is more susceptible to
 the external disturbance.

Another noteworthy result from this plot is that
the location of the minimum  depends on the vertical profile, see Table 1.
The minimum of the reduction factor occurs at 1.9 R$_D$ and 2.8 R$_D$  for a $sech^2$ and a $sech$ vertical distribution 
respectively. Thus, the disc starts to respond to the external potential from a smaller radius which is from within the optical radius, hence the stellar warps will occur. Recall that
the optical or stellar radius of a typical galaxy is observed to be $\sim$ 4-5 exponential disc scalelengths,
 beyond which the intensity decreases sharply (van der Kruit \& Searle 1982).
In contrast, note the much larger radius for the onset of warps of 5.1 R$_D$ for an  exponential disc,
close to the outer edge of the optical disc,
as also seen in Saha \& Jog (2006). Thus an exponential disc will mainly allow the warps to be seen only in HI which typically extends beyond the optical disc.

\begin{table}
\centering
  \begin{minipage}{140mm}
   \caption{Warp onset radius  
for different vertical profiles }
\begin{tabular}{lll}
Density profile, $\rho (z)$& R$_{min}$/R$_D$ & $({\delta}_1)_{min}$ \\%
$\rho (z)$(from eq.[2]) & & \\
\\
sech$^2$ & 1.9 & 0.56 \\
sech & 2.8 & 0.70 \\
exponential & 5.1 & 0.85 \\
\end{tabular}
\end{minipage}
\end{table}
This difference follows from the mathematical form of the response function and the fact that the disc gravity is mostly from matter close to the mid-plane for the $sech^2$ case. Hence the net contribution to the response potential (eq.[25])is a maximum at lower radius. 
The minimum value of the reduction factor is higher for $sech$ and the minimum is broader, this trend is even stronger for the exponential case. For the exponential case, the negative disc response due to the self-gravity of the disc is 
due to matter at larger z values and hence is less strongly dependent
on the radial mass distribution.
Hence the minimum in the plot of the reduction factor versus the radius is not pronounced, and the definition of R$_{min}$ is thus not so sharply defined.

 Fig. 1 thus shows that for a given perturbation potential, a stellar disc with a steeper vertical disc density distribution has a larger onset radius for warps, and
a stronger warp amplitude (due to a lower disc resistance as denoted by a higher $\delta_1$ value).
 Thus the stronger warps will have a steeper rise in the amplitude. 
This trend naturally explains the puzzling observation, namely
"the farther away a warp starts, the steeper it rises", which was noted by Sanchez-Saavedra
et al. (2003).

\subsection {Dependence on R$_{D}$ and $z_0$:}

For the $sech$ vertical distribution, we next plot the reduction factor $\delta_1$ vs. radius for different values of the disc scalelength R$_D$ for a constant vale of the disc scaleheight z$_0$ = 0.3 kpc (Fig. 2).
Similarly, in Fig. 3 we plot the reduction factor $\delta_1$ vs. radius for different choice of z$_0$
for a constant vale of the disc scalelength R$_D$ = 3 kpc. 
 For low R$_D$ (Fig. 2) and similarly for
low z$_0$ (Fig. 3) the self-gravity is lower hence the disc follows the
external perturbation well (higher $\delta_1$) and it starts at a small radius way inside the disc.
\begin{figure}
\centering
\includegraphics[width=3.4in,height=3.0in]{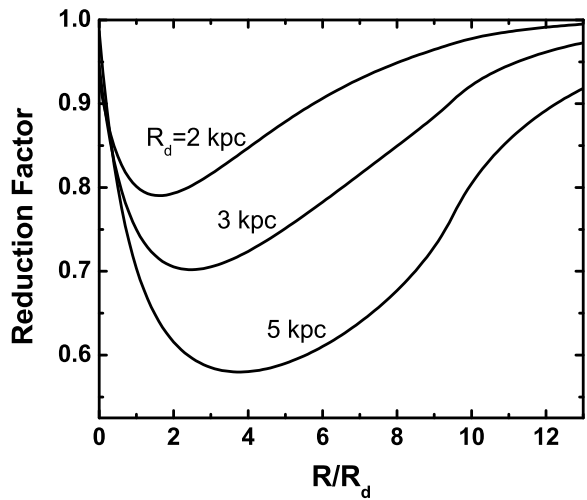}
\caption{Plot of $\delta_1$, the reduction factor, for $m=1$ vs. radius, R/ R$_D$, shown as a function of 
the scalelength $R_D$ for a given disc scaleheight $z_0$ = 0.3 kpc, for a vertical $sech$ density profile. The 
reduction factor is lower for a higher R$_D$ value since the disc is then more massive and hence shows a higher resistance to the external perturbation potential.}
\end{figure}
\begin{figure}
\centering
\includegraphics[width=2.8in,height=2.4in]{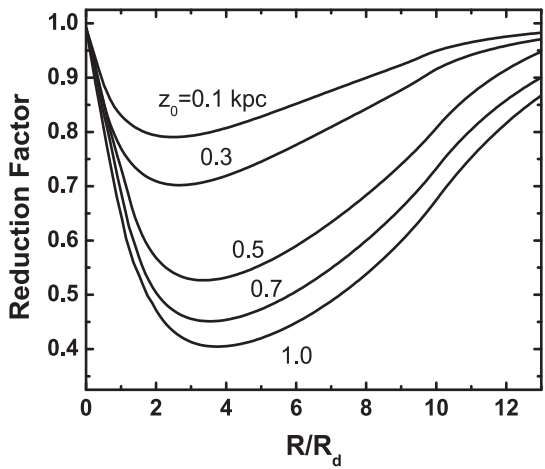}
\caption{Plot of $\delta_1$, the reduction factor for m=1 vs. the radius, R/R$_D$, shown as a function of 
the scaleheight $z_0$ for a given disc scalelength $R_D$ = 3 kpc, for a vertical sech density profile. The 
reduction factor is lower for a higher z$_0$ value since the disc is then more massive and hence shows a higher resistance to the external perturbation potential. }
\end{figure}

This dependence is further well-brought out  in a systematic study where the values for the input parameters are scanned, and the results obtained by solving eq.(28) in each case are given as contour diagrams, in the next two figures.
We vary the values for the disc scalelength, R$_D$ from 1 to 5 kpc and 
vary the vertical scaleheight z$_0$ from 0.1 to 1 kpc, while keeping the central density constant
at 1 M$_{\odot} pc^{-3}$ as before. This then scans the behaviour of galaxies of the same central density
but with varying mass - note that for a given central density, increasing R$_D$ and z$_0$ yields a more massive galaxy.
The resulting values of R$_{min}$, the radius where the reduction factor is the minimum for m=1, vs. radius are plotted in Figs. 4 a and 4 b for a vertical density distribution of type $sech$ (top panel) and exponential (lower panel) respectively.  In each case, the higher disc mass as given by higher R$_D$ and z$_0$ leads to a higher value of R$_{min}$ in the top r.h.s. corner of the plot,
hence the  warps set in at a higher radius. In contrast, for a lower mass galaxy 
(for the lower l.h.s. corner in the plot), the warps set in earlier. The results from this plot are general
and can be applied to any galaxy. Thus, for a typical giant spiral of a given central density, if its disc scalelength and the vertical scaleheight are known, then this figure gives the value of the minimum radius beyond which  warps can be seen. 
\begin{figure}
\centering
\includegraphics[width=3.2in,height=2.9in]{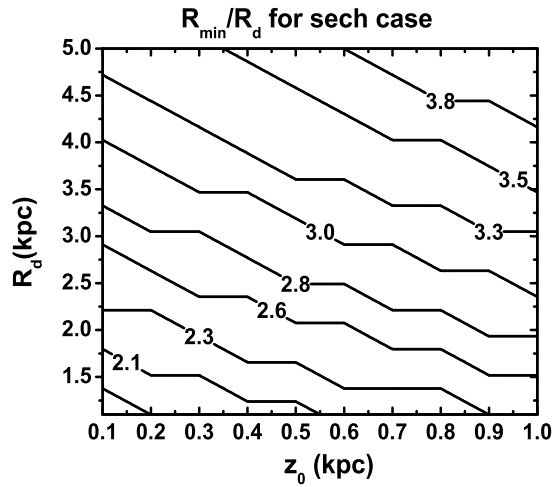}
\includegraphics[width=3.2in,height=2.9in]{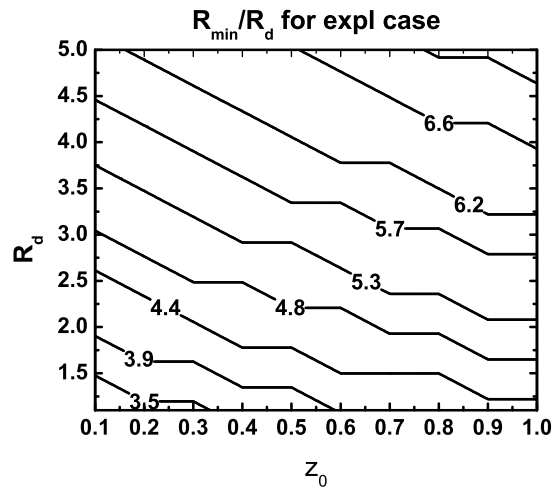} 
\caption{Contour plot of $R_{min}$ the radius at which the reduction factor is the minimum, shown as a function of 
 the disc scalelength $R_D$, and the scaleheight $z_0$; for the sech vertical density profile (top panel)
and the exponential vertical density profile (lower panel). The value is $<4$ for most parameters for the $sech$ case,
implying that stellar warps would always be seen in this case.}
\end{figure}

The two cases (Figs 4a and 4b) show a striking difference, namely that for the $sech$ profile for the vertical density distribution,  R$_{min}$ varies between 1.9 R$_D$ to 3.8 R$_D$. 
 Thus for all reasonable galactic parameters, the warps start to develop from within  the optical disc.
Hence the galaxies with a $sech$ profile or with $sech^{0.5}$ as observed, will {\it all} show
stellar warps. In contrast, for the steeper, exponential vertical density distribution
(lower panel), the R$_{min}$ covers a range of larger values, from 3.5 to 7.0 R$_D$. Thus over most of the parameter range, 
a disc with an exponential vertical profile cannot support stellar warps. This confirms the result in Saha \& Jog (2006).
Thus a seemingly small change in the
vertical density profile (a $sech$ vs. an exponential profile) has a drastic effect on the resulting radial onset of warps,
as already seen in Fig. 1.

Similarly, the results for the minimum of the reduction factor for $m = 1$ are plotted in Figs. 5a and 5b for a $sech$ 
vertical density profile (top panel) and for an exponential vertical density profile (lower panel). In each of Figs. 5a and 5b, we see that
for high z$_0$ and R$_D$ values (top r.h.s corner) where the disc is more massive, the reduction factor is the smallest meaning the negative disc response is the highest. Here the disc has the maximum resistance to being distorted. Whereas  at the lower l.h.s. corner, the disc is less massive, hence has less resistance due to self-gravity and hence responds more readily to the 
external perturbation. On comparing these two cases, one can see that the typical reduction factor has a smaller value
 for the $sech$ vertical profile. This is due to a greater overall resistance by the self-gravity of the disc to being distorted.
Thus the warps for the $sech$ case would be weaker 
and start at lower radii. Thus we predict that the resulting stellar warps are weaker compared to the HI warps,
this agrees with observations (Reshetnikov \& Combes 1998). 
\begin{figure}
\centering
\includegraphics[width=3.2in,height=3in]{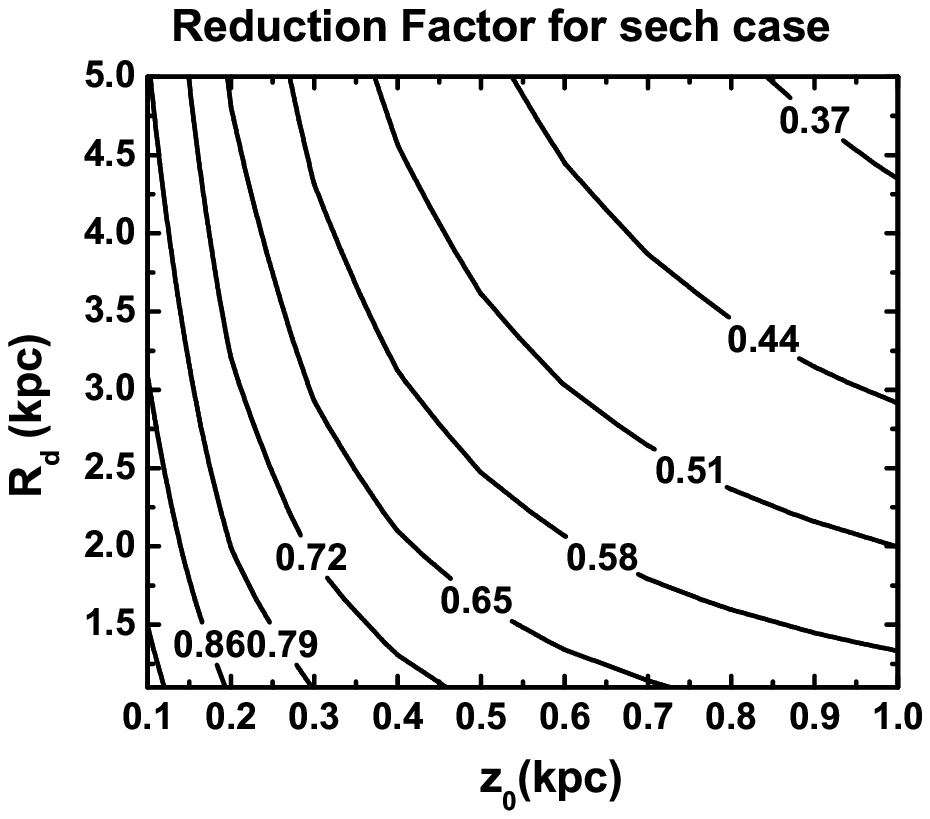}
\includegraphics[width=3.2in,height=3in]{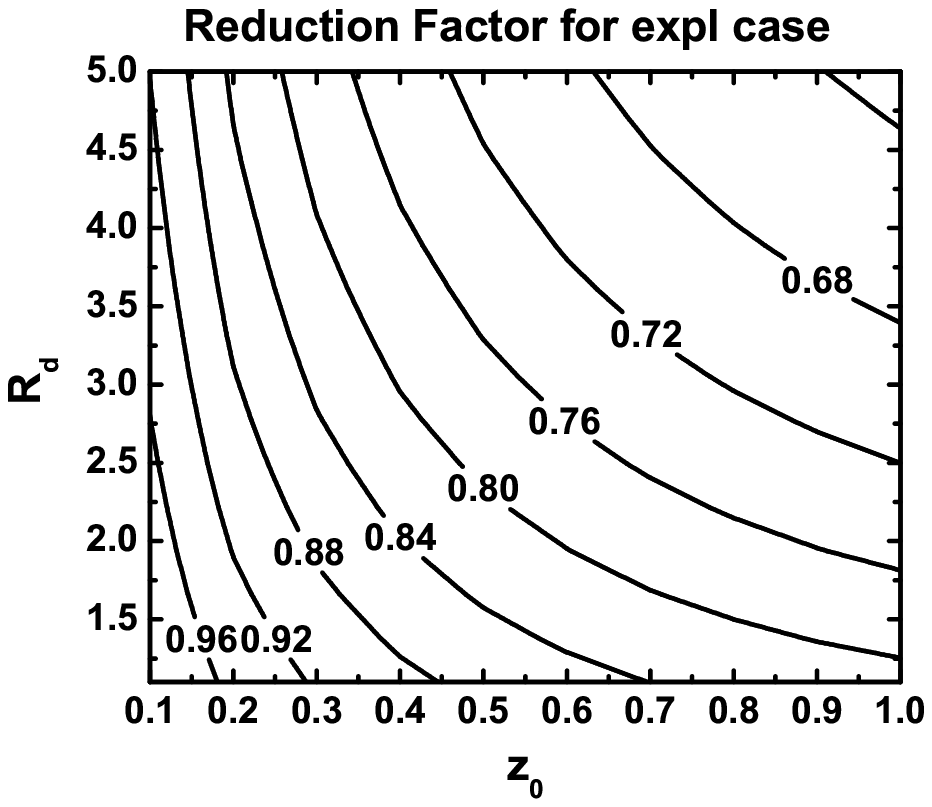} 
\caption{Contour plot of the minimum of the reduction factor, $\delta_1$, for $m=1$, shown as a function of 
the disc scalelength $R_D$, and the scaleheight $z_0$,
 for  a $sech$ vertical density profile (top panel) and for an exponential vertical density profile (lower panel). The reduction due to self-gravity is more significant for the $sech$ case.} 
\end{figure}

\subsection {Dependence on the Fourier component, $m$:}
The basis of the calculation in Section 2 is the fact that any function can be
represented as a sum of discrete Fourier components, where $m$ = 1,2... correspond to the
$m^{th}$ Fourier decomposition. It is but natural that when the density response
 is decomposed into the Fourier
components, the higher order terms will also have an overall contribution to
the final response potential. It is well-known that the higher order
perturbation terms show signatures in the disc too. To see the behaviour of these,
we calculate the reduction factor for a few select $m$ upto 10 for the $sech$ profile (Fig. 6).

A prominent saddle-shaped $m=2$ feature has also been seen for our Galaxy, in addition to the m=1 feature namely, the usual  integral-shaped warp (Levine, Blitz \& Heiles 2006). 
Figure 6 gives the
reduction factor, $\delta_2$ versus R/R$_D$ for this case. Similarly the higher order $m$ results are also shown for  $m= 3,5,$ and $10$, with an application in mind to the corrugations or scalloping observed in galaxies.
\begin{figure}
\centering
\includegraphics[width=2.8in,height=2.4in]{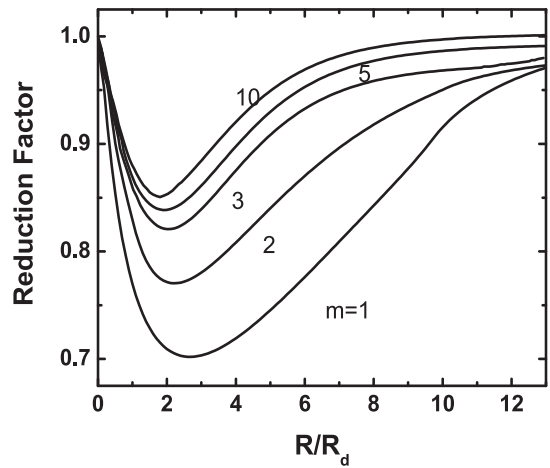}
\caption{Plot of the reduction factor $\delta_m$ for different Fourier component $m$ values vs. radius, R/ R$_D$, 
for a $sech$ vertical density distribution. Note that as $m$ increases, the values of the reduction factor are higher and the R$_{min}$ for the minimum of the reduction factor shifts to lower radii. Thus there is less resistance due to disc self-gravity to vertical
distortion. The disc is susceptible to higher $m$ modes from deep within the optical disc.}
\end{figure}
\begin{table}
\centering
  \begin{minipage}{140mm}
   \caption{Onset radius and reduction factor 
for vertical distortion,$m$ }
\begin{tabular}{lll}
$m$ component& R$_{min}$/R$_D$ & $({\delta}_m)_{min}$ \\
\\
1 & 2.8 & 0.70 \\
2 & 2.4 & 0.77 \\
3 & 2.1 & 0.82 \\
5 & 2.0 & 0.84 \\
10 & 1.9 & 0.85  \\
\end{tabular}
\end{minipage}
\end{table}
We note that the graph for $m=2$ lies higher compared to the graph
of $m=1$, and indeed all the higher $m$ cases show a  progressive trend (see Table 2).
After the onset, all the Fourier $m$ modes grow stronger with radius (Fig. 6), as is observed (Levine et al. 2006).
The reduction factor values including the minimum  for $m=2$ are %slightly 
higher than in the $m=1$ case. 
This could have
significant implications as it serves to tell us that $m=2$ signatures are
seen relatively universally in a galaxy. 
It is interesting that observationally for our Galaxy the various modes are indeed seen over a large radial range, starting from a few disc scalelengths
(Levine et al. 2006, Fig. 13). This is explained naturally by our result in Fig. 6.

Matthews  \& Uson (2008) have found a similar result for IC 2233, where they show that the ``corrugations'' in the disc, which
could correspond to high $m > 1$ signatures, are noticeable throughout the radial extent
of the galactic disc, which are stronger in the outer regions (see their Fig. 1).

The planar cases showed the same decreasing radial dependence of R$_{min}$ for higher $m$ components (Jog 2000). Saha \& Jog (2006) on the other hand got a higher R$_{min}$ for a higher m =10, they probably got this wrong since they had not taken the correct account of the radial dependence while solving for the self-consistent disc response numerically.

\section{Discussion}

\noindent {\it 1. Stellar warps and their detection:}

Observations show that stellar warps are common  and occur in more than 50\% of spiral galaxies(Sanchez-Saavedra, Battaner \& Florido 1990,
Reshetnikov \& Combes 1998). These therefore must start
within the Holmberg radius. Indeed this distinction though obvious is not often made in the literature- namely, a stellar warp by necessity must start within the optical radius.
  Recent systematic study of 325 edge-on galaxies Ann \& Park (2006) confirms this point, 
with a typical warp radius = 0.7 times the optical radius or $\sim 3$ disc scalelengths.
Interestingly, this observed value agrees well with our typical onset radius of $\sim 3$ disc scalelengths (Table 1). 
Thus, we have shown that a realistic, less-steep vertical density distribution of type $sech^{2/n}$ results in the onset of stellar warps within the optical radius as is observed in most galaxies.

 A similar value is seen for our Galaxy, where the stellar warp is shown to start from 3.1 disk scalelengths  based on the COBE/DIRBE data (Drimmel \& Spergel 2001), and 2.4 disk scalelengths based on the 2MASS data (Lopez-Corredoira et al. 2002).

The {\it Spitzer} observations of ten galaxies show the onset of warps lie within 3-6 disc scalelengths, thus many start from within the optical radius (Saha, de Jong \& Holwerda 2009).  These authors
 treat an exponential vertical density profile for simplicity, and try to explain the small observed warp onset radii by assuming that the scaleheight increases with radius - while keeping the disc mass constant. Such flaring with radius is observed in some galaxies (de Grijs \& Peletier 1997) and is expected for a multi-component, coupled, star-gas
 disc (Narayan \& Jog 2002).
However, the values of flaring they use are ad-hoc, and
 even this cannot explain the entire range of smaller values of R$_{min}$ that are observed for their sample. Further, they use $z_0/ R_D$ as a single thickness parameter but as we have shown (Fig. 4), the value of R$_{min}$ is not a simple function of this parameter. 
Instead the small onset radii
can be explained naturally as we have done, by using the observed less-steep vertical density distribution.

\medskip

\noindent {\it 2. Warp onset radius: Dependence on disc mass}

At high redshifts,
the galaxy size is smaller as seen for the Hubble deep field sample 
(Elmegreen et al. 2005) with a typical disc scalelength of 
1.5 kpc. 
Such a size variation with redshift is expected in the hierarchical 
evolution scenario (e.g., Steinmetz \& Navarro 2002). The warp onset radius for this sample is observed to be smaller $\sim$ 
1.4 disc scalelengths (Reshetnikov et al. 2002). This observed
trend agrees exactly  with our result (see Fig. 4)- namely that the smaller mass
galaxies allow warps to develop from a smaller starting radius.

At the opposite end, we predict that massive, nearby disc galaxies would have warp onset at a larger 
radius and thus are less likely to show a stellar warp. This would be tricky to confirm because the radial range over which optical warps are seen is small (starting at 3 R$_D$ and going up to 4-5 R$_D$), and observational 
data need to be analyzed to study this point.
    There is some evidence for this correlation: M31 has a massive 
disc with a scalelength of 5.4 kpc (Geehan et al. 2006),
and it has a stellar warp starting  at radii larger than 
the isophote at 26.8 ${\mu}_B$ (Innanen et al. 1982). This is beyond the Holmberg radius, 
whereas the average onset radius of stellar warps is within this radius, see point 1 above.

\medskip

\noindent {\it 3. Effect of nearby perturbers, and live halo :}

A real galaxy is likely to 
undergo 
close encounters with satellites less massive than the LMC, while our calculation is meant
for a distant encounter with distance 
large compared to the galaxy size.
    Further, for simplicity, we have taken the halo to be rigid and the 
forcing frequency to be zero. The long-standing problem about the tidal 
origin of warps has been the resulting small amplitude (Hunter \& Toomre 
1968). This is overcome if the halo is live and the wake generated at the 
resonance points of the frequency of the perturber's motion is included 
(Weinberg 1998, Tsuchiya 2002, Weinberg \& Blitz 2006).

     The Sagittarius dwarf is about 10 times less massive than the LMC but 
is three times closer, so their direct tidal torques on the Galaxy (being proportional 
to the mass of the perturber/distance$^3$ to the perturber) are comparable. 
Hence the Sagittarius dwarf also cannot directly produce the Galactic warp 
that is observed. However, as Bailin (2003) has argued, the magnitude of the tidal field due to 
the wake generated in the halo by the Sagittarius dwarf could also explain 
magnitude of the warp seen in the Galaxy. This needs to be checked by simulations. 

     Yet another pathway to create vertical perturbations would be to have
an even smaller mass perturber, like the subhalo, come even closer
to $\sim 5-10$ kpc. This possibility has been studied by Chakrabarti \& Blitz (2009) 
via simulations, who show that this can explain the HI amplitudes of warps 
and the higher order vertical modes as observed by Levine et al. (2006).

In the present paper we have not attempted to obtain the actual warp amplitude.
 However, the concept of negative disc response studied here would still 
apply in these general cases. That is, due to its strong self-gravity, the 
inner disc region would resist being distorted by vertical perturbations.

\medskip

\noindent {\it 4. Dynamical implications: } The present paper shows the surprisingly strong dependence of the resulting warp radius on the
vertical density distribution in the disc. This is because the disc self-gravity which decides
the warp radius depends crucially on the vertical disc distribution close to the mid-plane.
A similar strong dependence on the disc vertical distribution may also affect other dynamical studies
such as the vertical heating due to tidal encounters, or the bending instabilities. We plan to look at these in future papers.

\section {Conclusions}

We study the self-consistent, linear response of a galactic disc to vertical perturbations arising  due to a tidal encounter. 
We show that the vertical disc density distribution has a surprisingly strong effect on the radius at which disc starts to show the vertical distortions. In retrospect this is physically understandable since the disc resists distortion in the inner parts due to its self-gravity, which in turn depends  crucially on the vertical mass distribution close to the mid-plane. A flat-core profile like the $sech$ as is observed results in a small radius for onset of warps of $\sim$ 3 disc scalelengths. We can thus naturally explain why the radius of onset of stellar warps is within the optical disc as observed in most galaxies. 
The results for the radius for warp onset are given as a function of the disc scalelength and disc scaleheight, and are presented as contour plots, which can therefore be applied to any galaxy.

We show that the higher $m$ distortions develop from even smaller radii $\sim 2$ disc scalelengths. Hence the higher mode corrugations can be seen over a large radial range, in agreement with what is observed for the Galaxy
(Levine et al. 2006), and for IC2233 (Matthews \& Uson 2008).

\bigskip

\medskip

\noindent {\bf Acknowledgments:}  $\: \: $ We would like to thank the anonymous referee for 
insightful comments, especially for asking us to include the effect of the shape of the dark matter halo on the warp.
 We also thank Lynn Matthews for useful comments.

\bigskip

\bigskip

\noindent {\bf References}

\medskip

\noindent Ann H.B., Park J.-C., 2006, New Astronomy, 11, 293

\noindent Bailin J., 2003, ApJ, 583, L79

\noindent Bailin J., Steinmetz M., 2005, ApJ, 627, 647

\noindent  Banerjee A., Matthews L.,  Jog C.J., 2010, New A, 15, 89

\noindent Banerjee A.,  Jog C.J., 2008, ApJ, 685, 284

\noindent Banerjee A.,  Jog C.J., 2007, ApJ, 662, 335

\noindent Barteldrees A., Dettmar R.-J., 1994, A \& A S, 103, 475

\noindent Bett  P., Eke  V., Frenk  C.S., Jenkins A., Helly  J.,  Navarro J.F.,
2007, MNRAS, 376, 215

\noindent Briggs F., 1990, ApJ, 352, 15

\noindent Binney J., Tremaine S., 1987, Galactic Dynamics. Princeton
Univ. Press, Princeton, NJ

\noindent Chakrabarti S., Blitz L., 2009, MNRAS, 399, L118

\noindent de Grijs R., Peletier R.F., 
  1997, A \& A, 320, L21

\noindent de Grijs  R., Peletier  R.F.,  van der Kruit P.C. 
  1997, A \& A, 327, 966

\noindent de Zeeuw T., Pfenniger D., 1988, MNRAS, 235, 949

\noindent Drimmel R. , Spergel D.N., 2001, ApJ, 556, 181

\noindent  Elmegreen B.G., Elmegreen D.M., Vollbach D.R., Foster E.R.,  Ferguson T.E., 2005, ApJ, 634, 101

\noindent Freeman K.C., 1970, ApJ, 160, 811

\noindent Florido E., Battaner E., Prieto M., Mediavilla E.,  Sanchez-Saavedra, M.N.,
   1991, MNRAS, 251, 193

\noindent Geehan J.J., Fardal M.A., Babul A., Guhathakurta  P.,  2006, MNRAS, 366, 996

\noindent Gilmore G., Reid N., 1983, MNRAS, 202, 1025

\noindent  Hunter C., Toomre A., 1969, ApJ, 155, 747

\noindent Innanen K.A., Kamper K.W., Papp K.A.,  van den Bergh S., 1982,  ApJ, 254, 515

\noindent Jackson J.D., 1975, Classical Electrodynamics, Wiley, New York

\noindent Jog C.J., 1999, ApJ, 522, 661

\noindent Jog C.J., 2000, ApJ, 542, 216

\noindent Jog C.J., Combes F., 2009, Physics Reports, 471, 75

\noindent Kalberla P.M.W., Dedes L., Kerp J.,  Haud U., 2007, A \& A, 469, 511

\noindent Kent S.M., Dame T.M., Fazio G., 1991, ApJ, 378, 131

\noindent Kuhlen M., Diemand J., Madau P., 2007, ApJ, 671, 1135

\noindent Kuijken K., Gilmore G., 1989, MNRAS, 239, 571

\noindent Kulkarni S.R., Blitz L.,  Heiles C., 1982, ApJ, 259, L 63

\noindent Levine E.S., Blitz L., Heiles C., 2006,  ApJ, 643, 881

\noindent Lopez-Corredoira M., Cabrera-Lavers A., Garzon F., Hammersley P.L., 2002, A \& A, 394, 883

\noindent Matthews L.D, Uson J., 2008, ApJ, 688, 237

\noindent Mera D., Chabrier G.,  Schaeffer R., 1998, A \& A, 330, 953

\noindent Narayan C.A.,  Jog  C.J., 2002, A \& A, 390, L35

\noindent Narayan C.A., Jog  C.J., 2002, A \& A, 394, 89

\noindent Narayan C.A., Saha K., Jog C.J., 2005, A\&A, 440, 523

\noindent Quiroga R.J., 1974, Ap\&SS, 27, 323

\noindent Reshetnikov V., Battaner E., Combes F.,  Jimenez-Vicente J., 2002, A \& A, 382, 513

\noindent Reshetnikov V.,  Combes F., 1998, A \& A, 337, 9

\noindent Rice W., Merrill K.A., Gatley I.,  Gillett F.C., 1996, AJ, 112, 114

\noindent Sackett P.D.,  Sparke L.S., 1990, ApJ, 361, 408

\noindent Saha K., Jog C.J., 2006, MNRAS, 367, 1297

\noindent Saha K., de Jong R., Holwerda B. 2009, MNRAS, 396, 409

\noindent Sanchez-Saavedra M.L., Battaner E., Guijarro A., Lopez-Corredoira M.,  Castro-Rodriguez N., 2003,
 A \& A, 399, 457

\noindent Sanchez-Saavedra M.L., Battaner E., Florido E., 1990, MNRAS, 246, 458

\noindent Schwarz U.J., 1985, A \& A, 142, 273

\noindent Spitzer L., 1942, ApJ, 95, 329

\noindent  Steinmetz M.,  Navarro J.F., 2002, New Astronomy, 7, 
155

\noindent Tsuchiya T., 2002, New A, 7, 293

\noindent van der Kruit P., 1988, A \& A, 192, 117

\noindent van der Kruit P., Searle L., 1982, A \& A, 110, 66

\noindent Wainscoat R.J., Freeman K.C.,  Hyland A.R.,    1989, ApJ, 337, 163

\noindent Weinberg M.D.,  Blitz L., 2006, ApJ, 641, L 33

\noindent Weinberg M.D., 1998, ApJ, 299, 499

\noindent Weinberg M.D., 1995, ApJ, 455, L31

\noindent Zaritsky D., Rix H.-W., 1997, ApJ, 477, 118 

\bigskip

\bigskip

\bigskip

\noindent {\bf Appendix A:  Weak dependence of warp on the shape of the dark matter halo}

\medskip

We have included a spherical dark matter halo in the calculations (Sections 2 and 3)
for the sake of completeness. We show next that it has a negligible effect on the
vertical response of the disk including the warp onset radius.
In this Appendix, we also investigate the possible effect of the shape of the dark matter halo
on the warp. Cosmological N-body simulations show that the dark matter halos are not spherical 
but instead tend to be flattened, with the typical value of $q = c/a$,  the vertical 
to planar axes ratio of 0.6 (Bailin \& Steinmetz 2005) or  0.7 
(Bett et al. 2007; Kuhlen, Diemand \& Madau 2007).

     In this paper, we have studied the self-consistent response of a galactic disc 
to an external perturbation, and the halo affects this calculation in an indirect way. 
The external, tidal perturbation 
affects both the disc and the halo. 
The self-consistent disc response is obtained by treating its response both to the external potential,
plus the response potential corresponding to the disc and halo response 
to the net perturbation (eq. [34]).
Thus the inclusion of halo enters in  the calculation of the frequencies (eq. [14]),
and the net response density (eq. [18]). Of these, the frequencies ${\nu}^2$ 
and ${\Omega}^2$ are affected by the halo, especially in the outer parts and for an oblate halo.
On the other hand, the response density of the halo is 
small $\sim 10^{-3} \times$ the disc response density, hence it does not affect the self-consistent disc response. This is because the halo, even an oblate one, 
is much less concentrated towards the disc mid-plane. 

We solve eq.(28) (see Section 3.1) and obtain the reduction factor $\delta_1 $ vs. radius for a sech vertical profile of the disc, for a disc-alone case, and the disc plus halo cases for $q$ = 0.95 (spherical case), q=0.7 (the typical value seen in cosmological simulations), and q=0.4 (an extremely flattened oblate halo). 
The disc and the halo parameters
for the spherical case are as given in Section 3.1. 
For the flattened halo cases the relation between the central density $\rho_c$
and the core radius $R_c$ are obtained in terms of those for a spherical halo, assuming a constant halo mass, and using the relations derived in Narayan et al. (2005, see their eq. [7]).
The results are plotted in Fig. 7.

\begin{figure}
\centering
\includegraphics[width=2.8in,height=2.4in]{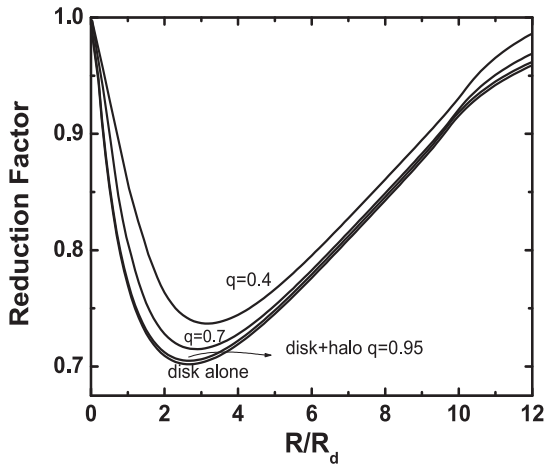}
\caption {Plot of the reduction factor $\delta_1$ for disc-alone, and for disc-plus-halo with different oblateness of the halo , with $q$ = 0.95 (spherical) to q=0.4 (highly flattened halo) vs. radius, R/ R$_D$, 
for a $sech$ vertical density distribution. The spherical halo has a negligible effect on the net disc response. Even a flattened halo with the axis ratio, q =0.7 as seen in cosmological simulations, plays a minor role in determining the disk response, changing the radius of onset of warps by just $ 10 \%$.}

\end{figure}

The most striking result is that the inclusion of a spherical halo has a negligible effect on the
disc response. The lowest two curves are nearly identical.
The weak dependence on the halo may seem surprising since the halo
is known to be dominant in the outer parts. However, there are two reasons 
for this: first, the disc mass and the halo mass are comparable upto the 
region of interest, namely R$_{min}$ ; second, the halo is much less 
concentrated towards the disc mid-plane.
 Thus, the inclusion of a spherical halo has a negligible effect on the radius of onset of warps given by the location of the minimum in the reduction factor curve  (see Table 3).

\begin{table}
\centering
  \begin{minipage}{140mm}
\caption{Warp onset radius  
for different halo flattening }
\begin{tabular}{lll}
Component& R$_{min}$/R$_D$ & $({\delta}_1)_{min}$ \\
\\
disc-alone & 2.6 & 0.70 \\
disc-plus-halo (q=0.95, spherical) & 2.7 & 0.71 \\
disc-plus-halo (q=0.7, flattened)& 2.9 & 0.72 \\
disc-plus-halo (q=0.4, very flattened) & 3.2 & 0.74 \\ 
\end{tabular}
\end{minipage}
\end{table}

The shape of the halo is also shown to play a minor role in determining the disc response.
A typical oblate halo, with a flattening of $q=0.7$ as seen in the cosmological simulations. 
changes the radius of onset of warps by just $ 10 \%$ (Table 3). 
Interestingly, even a highly flattened halo with $q=0.4$, which occurs at the most oblate end of
the halo distribution found in simulations (e.g., Bett et al. 2007, Bailin \& Steinmetz 2005), has a small effect on the radius of warps. Thus over the entire range of flattening of the halos seen in cosmological simulations, the halo
has a minor effect on the determination of the warp.

\end{document}